\documentclass[aps,twocolumn,amsmath,amssymb,10pt,floatfix]{revtex4-1}
\usepackage{graphicx,float}

\begin{document}
\title{Analytical solutions of the two-dimensional Dirac equation for a topological channel intersection}

\author{J.R. Anglin and A. Schulz}

\affiliation{\mbox{State Research Center OPTIMAS and Fachbereich Physik,} \mbox{Technische Univerit\"at Kaiserslautern,} \mbox{D-67663 Kaiserslautern, Germany}}

\begin{abstract}
Numerical simulations in a tight-binding model have shown that an intersection of topologically protected one-dimensional chiral channels can function as a beam splitter for non-interacting fermions on a two-dimensional lattice \cite{Qiao2011,Qiao2014}. Here we confirm this result analytically in the corresponding continuum $\mathbf{k}\cdot\mathbf{p}$ model, by solving the associated two-dimensional Dirac equation, in the presence of a `checkerboard' potential that provides a right-angled intersection between two zero-line modes. The method by which we obtain our analytical solutions is systematic and potentially generalizable to similar problems involving intersections of one-dimensional systems.
\end{abstract}
\date{\today}

\maketitle

\section{Introduction}
Two-dimensional Dirac fermions, such as are realized to a good approximation in many effectively two-dimensional systems,
can possess topologically protected edge-state modes which are effectively one-dimensional. As well as propagating along the literal edges of finite two-dimensional systems, these modes can form along zero-potential lines within a bulk sample, for instance of graphene \cite{Jung2011,Martin2008,Ju2015,Li2016}. Such zero-line modes (ZLMs) may thus potentially serve as nano-electronic `wires made of nothing' that can be written into two-dimensional substrates by means of externally applied potentials. To apply this concept in non-trivial circuits with junctions, however, will further require a means of effectively `soldering' two such `wires' together in a way that will allow their chiral currents to split or combine.

Intersecting ZLMs have recently been studied numerically \cite{Qiao2011,Qiao2014}, in a non-interacting lattice model that represents the $\pi$-band electrons of graphene with a position-dependent sublattice-staggered external potential constructed to form intersecting zero lines (with suggestions and references for other physical realizations of models with similar zero-line intersections). Because of the chiral nature of the ZLMs, two of the half-lines which meet at the intersection support only incoming modes, while the other two support only outgoing modes. The numerical results for the Landauer-B\"uttiker conductances across the intersection confirm that chiral currents can indeed split (current in one incoming ZLM exits in both outgoing modes) or combine (the opposite). Figure~1 shows a sketch of the qualitative effect.

\begin{figure}
\center\includegraphics[width=1\columnwidth]{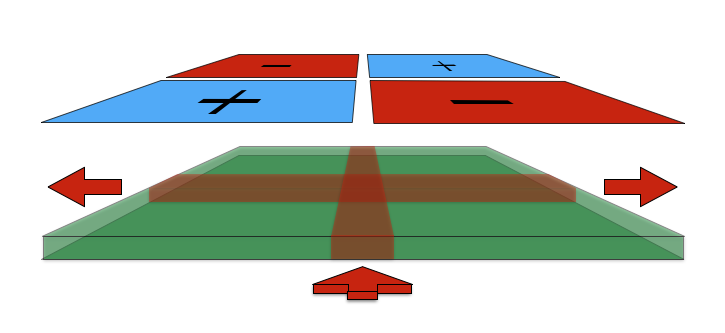}
\caption{Alternating positive and negative electrodes (above) induce a ``checkerboard''
potential landscape within a sheet of graphene (green sheet). Topologically protected zero-line modes (ZLMs) propagate one-dimensionally along the intersecting lines of zero potential (red stripes). Particles in an incoming ZLM (indicated by the
bottom red arrow) neither cross the intersection directly ahead, nor
reflect backwards; they split in equal coherent superposition into both left and right outgoing ZLMs (shown by outgoing arrows).}
\end{figure}

Here we support and extend these specific numerical results, by analytically solving
the two-dimensional Dirac problem for the associated $\mathbf{k}\cdot\mathbf{p}$ model. Although \cite{Qiao2011,Qiao2014} have found interesting results for intersections at different angles, we consider here only the simplest case of a right-angled intersection. Our solution is obtained as an integral representation, in which the exact Dirac energy eigenspinors for all energies within the bulk gap are given by a one-dimensional integral whose integrand we obtain explicitly. This integral is in fact quite complicated, but it allows analytic determination of asymptotic behaviors, including the numerically observed splitting, which we confirm to be coherent. If necessary, the detailed form of the eigenspinors near the intersection itself can also be obtained by straightforward numerical integration. The method by which we obtain our exact integral representations is systematic, and potentially generalizable to other problems with piecewise continuous potentials in two dimensions.

Plots of our solutions show that they are simple in form despite their complex representation. In particular, we see that chiral particles occupying the ZLM turn the corner quite sharply at the intersection, with little spreading away from the zero-lines, even at energies near the edge of the bulk gap. Although one might expect that the abrupt acceleration of a 90-degree corner would tend to loosen the particles from the ZLM's one-dimensional grip, our results show that this essentially does not occur. The tightness of ZLM binding to the zero-lines, even through the intersection, helps to explain the robust retention of mode chirality that was observed in the discrete numerical computations of \cite{Qiao2011}. 

Our paper is structured as follows. We first review the basic concepts of Dirac ZLMs in Section II, 
and then in Section III we present the Dirac problem for the topological intersection, and also show the form of analytical solution that we will find. In Section IV we show how a continuous global solution to the Dirac equation can be composed by adding together a certain set of many local solutions that are only piecewise continuous, as long as the weights with which these local solutions are added satisfy a certain integral equation. In Section V we then solve this integral equation for all energy eigenvalues within the bulk gap, using a systematic method that can in principle be extended to other problems, to produce an analytic solution to the Dirac problem in the form of an integral representation. In Section VI we collect and summarize our analytical results, and plot our eigenspinors as numerical integrations of the exact integral representations. We conclude in Section VII with discussions of possible directions for future work. Our discussion of extensions to interacting systems includes an explanation of how to represent the topological intersection one-dimensionally as a kind of impurity that connects two one-dimensional lines.

\section{Zero-line modes}

Our analysis in this paper will be based on a continuum theory of non-interacting two-dimensional fermions with momenta near a Dirac point, with a position-dependent Dirac mass. This theory may be derived from a nearest-neighbor tight-binding model in the presence of an appropriate external potential \cite{Qiao2014}, as long as the potential varies slowly on the lattice scale, by applying the lowest-order $\mathbf{k}\cdot\mathbf{p}$ approximation \cite{Martin2008}. In appropriate units, the low-energy Hamiltonian for a single particle near the `K'  Dirac point then takes the relativistic form
\begin{eqnarray}\label{Ham0}
\hat{\mathcal{H}}=\hat{p}_{x}\sigma_{x}+\hat{p}_{y}\sigma_{y}+U(x,y)\sigma_{z}
\end{eqnarray}
where the momentum operators are $\hat{p}_{j}=-i\partial_{j}$, and the
Pauli matrices $\sigma_{i=x,y,z}$ act on the sub-lattice pseudo-spin. The time-independent Schr\"odinger equation for $\hat{\mathcal{H}}$ is then the time-independent Dirac equation
\begin{eqnarray}\label{Dirac}
E\,\Psi = \hat{\mathcal{H}}\Psi
\end{eqnarray}
for 2-spinors $\Psi(x,y,E)$. Since $\Psi^{*}$ then obeys (\ref{Dirac}) except with $\hat{p}_{x}\to -\hat{p}_{x}$, solutions for the opposite K$'$ Dirac point can be obtained trivially from the solutions we present.

The zero-line modes appear most simply when the continuum potential has the form of a step across the $x$-axis: $U(x,y)\to U\,\mathrm{sgn}(y)$ for constant $U$. It is then easy to see that while eignspinors which extend over the entire $x,y$ plane all have energies $|E|\geq |U|$, eigenspinor solutions $\Psi\to W_{x}(x,y,E)$ with $|E|<|U|$ are given by
\begin{eqnarray*}
 W_{x}(x,y,E)& = & e^{-|U|\,|y|}e^{-i\,\mathrm{sgn}(U) Ex} \left(\begin{matrix}1\\ -\,\mathrm{sgn}(U)\end{matrix}\right)\;.
\end{eqnarray*}
These solutions that exist within the bulk energy gap $|E|<|U|$  are all confined to within a distance $\sim 1/|U|$ from the potential zero line $y=0$ (the $x$-axis). The one-dimensionally confined subgap modes $W_{x}$ exist in only one spin state, polarized along the $x$-axis with direction determined by the sign of the potential $U$; the direction of spin is also the direction of propagation. These quasi-one-dimensional chiral modes are simple extensions to two dimensions of the isolated zero-energy bound states that are found in one-dimensional Dirac systems with topologically non-trivial boundary conditions \cite{Jackiw1976,Su1979}.

Alternatively we could consider a $y$-axis zero line, $U(x,y)\to U\mathrm{sgn}(x)$. In this case we would find the slightly different eigenspinors
\begin{eqnarray*}
 W_{y}(x,y,E)& = & e^{-|U|\,|x|}e^{i\, \mathrm{sgn}(U) Ey} \left(\begin{matrix}1-i\,\mathrm{sgn}(U)\\ 1+i\,\mathrm{sgn}(U)\end{matrix}\right)\;.
\end{eqnarray*}
Again the one-dimensional states can have only one spin orientation, and it is in the direction of their motion.

These examples of abruptly reversing potentials are clearly not compatible with the condition that $U$ vary slowly on the lattice scale. In fact, however, the existence and effectively one-dimensional nature of ZLMs are not sensitive to the details of how the potential varies around the zero line, because they are topologically necessary \cite{Qi2006}. Indeed, studies of ZLMs in gated bilayer graphene with a smoothly varying potential have found little deviation of important properties in the regime we consider \cite{Martinez2009,Zarenia2011}. Hence the step potential form for $U(x,y)$ may be considered as a conveniently solvable idealization, within the continuum theory, of a real potential which varies more slowly, and for which the continuum theory is valid. 

In the following Section we will consider the more complex `checkerboard' potential $U(x,y) = U\,\mathrm{sgn}(x)\,\mathrm{sgn}(y)$ that defines the \emph{topological intersection}. It is also to be considered as a topologically equivalent idealization of a smoother potential to which the continuum theory applies. Numerical calculations for the tight-binding model have confirmed that the continuum eigenspinors which we present in the next Section do indeed correspond to eigenstates in the tight-binding theory, when the idealized potential is made more realistically smooth, so that particles are never scattered out of the $\mathbf{k}$-space neighborhood of the K Dirac point which is represented by the continuum theory.

\section{Intersecting zero-line modes}

\subsection{Simplifications by symmetry}

First of all, we can avoid duplication of effort by noting certain symmetries of the Dirac equation with $U(x,y) = U \mathrm{sgn}(x)\mathrm{sgn}(y)$
\begin{eqnarray}\label{Dirac1}
E \Psi = U\mathrm{sgn}(x)\mathrm{sgn}(y)\sigma_{z}\Psi -i\left[ \sigma_{x}\partial_{x}+\sigma_{y}\partial_{y}\right]\Psi\;.
\end{eqnarray}
We can consider $U>0$ without loss of generality, because the case of negative $U$ can be obtained from it by substituting $\Psi\to\sigma_{z}\Psi$ and $E\to -E$. Then we can restrict our attention to $E>0$ as well, because for every 2-spinor $\Psi(x,y)$ satisfying $\hat{H}\Psi = E\Psi$, the conjugate $\Psi_{T} =\sigma_{x}\Psi^{*}$ satisfies $\hat{H}\Psi_{T} = -E\Psi_{T}$. 

Equation (\ref{Dirac1}) also has three reflection symmetries: for every 2-spinor $\Psi(x,y)$ satisfying $\hat{H}\Psi = E\Psi$, both $\Psi_{R1} =\sigma_{x}\Psi(x,-y)$ and $\Psi_{R2} = \sigma_{y}\Psi(-x,y)$ satisfy the same equation --- as does the image of both transformations in succession, $\Psi_{R3} = \sigma_{z}\Psi(-x,-y)$. Our energy eigenvalues $E$ will not be four-fold or eight-fold degenerate, however, but only two-fold, because these three reflection operations clearly do not commute with each other. In fact they form an SU(2) group, of which our doubly degenerate eigenfunctions form a spin-1/2 representation. 

Without loss of generality we will be able to find energy eigenspinors $\Psi\to\Psi_{E}(x,y)$ which fulfill (\ref{Dirac1}) and are also even eigenstates of the $y$-parity operation $R_{1}$: $\sigma_{x}\Psi_{E}(x,-y)=\Psi_{E}(x,y)$. We can then obtain the orthogonal solution with the same $E$, $\tilde{\Psi}_{E}=i\sigma_{y}\Psi_{E}(-x,y)$. This means that $\tilde{\Psi}_{E}$ is  an odd eigenstate of $R_{1}$:
\begin{eqnarray}
	\sigma_{x}\tilde{\Psi}_{E}(x,-y) &\equiv& i\sigma_{x}\sigma_{y}\Psi_{E}(-x,-y)\nonumber\\
	&\equiv& -i\sigma_{y}\sigma_{x}\Psi_{E}(-x,-y)\nonumber\\
&\equiv&-i\sigma_{y}\Psi_{E}(-x,y)\equiv -\tilde{\Psi}_{E}\;.
\end{eqnarray}
These two solutions will therefore be orthogonal even though they are degenerate, because for them the Dirac inner product associated with (\ref{Dirac1}) reads
\begin{eqnarray}\label{DIP}
	\langle\tilde{\Psi}_{E}|\Psi_{E}\rangle &=& -i\int dxdy\,\tilde{\Psi}_{E}^{\dagger}(x,y)\Psi_{E}(x,y)\nonumber\\
&=&  i\int dxdy\,\tilde{\Psi}_{E}^{\dagger}(x,-y)\sigma_{x}\sigma_{x}\Psi(x,-y)_{E}\nonumber\\
& \equiv& - \langle\tilde{\Psi}_{E}|\Psi_{E}\rangle
\end{eqnarray}
since $\sigma_{x}^{2}\equiv 1$ and $y\to -y$ is a dummy variable. Arbitrary linear combinations of $\Psi_{E}$ and $\tilde{\Psi}_{E}$ are of course possible, and from these we may construct alternative solution pairs which are eigenstates of any of the other parity operations $R_{2}$ or $R_{3}$. Our choice to make eigenstates of $R_{1}$ was thus arbitrary.

The set of solution pairs $\Psi_{E}$ and $\tilde{\Psi}_{E}$ for $|E|<U$ span the whole space of ZLMs, because our intersecting axes represent two zero potential lines, and so we must expect twice as many independent solutions as would be present with a single zero line. These solutions are not complete in the sense that they provide a partition of unity in the full Dirac Hilbert space: for that, one must also include all the $|E|>U$ solutions, which propagate throughout the $(x,y)$ plane. Indeed the $\Psi_{E}$ and $\tilde{\Psi}_{E}$ solutions do not even provide a partition of unity within the subspace of Dirac excitations on the $x$ and $y$ axes, because each axis supports only one spin state as a ZLM. Excitations which pass through $x=0$ or $y=0$ with opposite spin to the corresponding ZLM must be composed of two-dimensional excitations with $|E|>U$.

\subsection{Form of the exact solution for $\Psi_{E}$}

So that the reader can keep our ultimate destination in mind while advancing through the details of our derivation, we begin by presenting our ultimate result in compact form.
From the exact integral representation to be derived below, we will be able to see that $\Psi_{E}$ takes the general form
\begin{widetext}
\begin{eqnarray}\label{psiE}
\Psi_{E}(x,y) &=& \theta(-x) e^{i[Ex+\alpha(E/U)]}e^{-U|y|}\left(\begin{matrix}1\\ 1\end{matrix}\right)
+\frac{1}{2}e^{-U|x|}e^{i[E|y|-\alpha(E/U)]}\left[\left(\begin{matrix}1\\ 1\end{matrix}\right)-i\,\mathrm{sgn}(y)\left(\begin{matrix}1\\ -1\end{matrix}\right)\right]+\Delta\Psi_{E}(x,y)
\end{eqnarray}
\end{widetext}
where $\theta(-x) = 1$ for $x<0$ and 0 for $x>0$; the phase shift $\alpha(E/U)$ is a certain function that we will later define exactly; and the origin-region correction $\Delta\Psi_{E}(x,y)$ is localized near $x=y=0$, decaying exponentially with $\sqrt{x^{2}+y^{2}}$. $\Delta\Psi_{E}$ incorporates discontinuities such that the entire wave function $\Psi_{E}$ is continuous everywhere. To describe this $\Psi_{E}$ in words, it consists of an incoming wave along the negative $x$-axis, with spin polarized in the positive $x$ direction, which splits equally into two outgoing waves along the positive and negative halves of the $y$-axis, each having spin polarized in their directions of motion (\textit{i.e.} in the positive and negative $y$ directions, respectively). 

There is no amplitude to propagate outward along the positive $x$-axis; the topological intersection acts as a symmetrical T-junction. The corresponding $\tilde{\Psi}_{E}=i\sigma_{y}\Psi_{E}(-x,y)$ has an incoming wave along the positive $x$-axis and no transmission along the negative $x$-axis. The linear combinations $\Psi_{E}\pm\tilde{\Psi}_{E}$ provide the eigenfunctions of $R_{2}$, which instead lack support along either the positive or negative $y$-axis, having an outgoing wave along the other half of the $y$-axis, and incoming waves on both halves of the $x$-axis. There are no solutions to (\ref{Dirac1}) with incoming waves on the $y$-axis or outgoing ones on the $x$-axis; one has such solutions, instead of the ones presented here, for the different Dirac equation with $\hat{p}_{x}\to -\hat{p}_{x}$ that applies near the K$'$ Dirac point. 

The topological intersection can thus both split and combine incoming ZLMs, coherently: it behaves exactly as a 50/50 optical beamsplitter, but for chiral fermions instead of photons.

In the low-energy limit $E\ll U$, we can be more explicitly quantitative. The phase shift $\alpha(E/U)$ satisfies $\alpha(\epsilon)=2\epsilon/\pi + \mathcal{O}(\epsilon^{3})$ for small argument, and this linear behavior of $\alpha$ at low energies implies that low-energy wave packets will emerge from the origin advanced by the distance $4/(\pi U)$, in comparison with the position one would expect if they followed the axes around sharp 90$^{\circ}$ corners at the origin; in effect the wave packets use their finite extension into the two-dimensional plane to `cut' the corner slightly. 

We can also show that $\Delta\Psi_{E\ll U}(\mathbf{r}) \sim e^{-U|\mathbf{r}|}$ is localized within a distance of order $U^{-1}$ from the origin, and hence further recognize that $\Delta\Psi_{E}\doteq\Delta\Psi_{0}$, because corrections proportional to $E\ll U$ have negligible effect on a function whose support does not extend to distances $|\mathbf{r}|\sim E^{-1}$. In contrast the asymptotically propagating parts of $\Psi_{E}$ do depend significantly on $E$ even when $E$ is small, since small differences in $E$ make large changes in $Ex$ or $Ey$ when $|x|$ or $|y|$ are large.

We now proceed to construct our integral representation solution to the checkerboard Dirac problem.

\section{Local solutions in half-planes}
We will solve the time-independent Dirac equation
\begin{eqnarray}\label{TDD}
E \Psi = \left[U\mathrm{sgn}(x)\mathrm{sgn}(y)\sigma_{z} -i( \sigma_{x}\partial_{x}+\sigma_{y}\partial_{y})\right]\Psi
\end{eqnarray}
for $U>0$ and $0\leq E<U$. We use $\Psi\to\Psi_{E}(x,y)$ to denote the full exact solution to this equation, continuous everywhere; we will build up this full solution $\Psi_{E}$ out of several component solutions, distinguished with various subscripts, that solve the equation locally, but are not everywhere continuous separately. As a first-order differential equation, the Dirac equation (\ref{TDD}) requires a continuous solution, though it need not be everywhere continuously differentiable.
The main task of our derivation will be to combine our discontinuous pieces in such a way that their discontinuities all cancel.

A local ZLM solution that is familiar from the simpler problem with a potential $V(x,y)=-U\mathrm{sgn}(y)$ is
\begin{eqnarray}\Psi_{E-} = \left\{
\begin{matrix}e^{iE x}e^{-U|y|}\left(\begin{matrix} 1\\ 1\end{matrix}\right)&,\; x<0\\ 0 &,\; x>0 \end{matrix}\right.\;.
\end{eqnarray}
$\Psi\to\Psi_{E-}$ solves our more complicated intersection problem (\ref{TDD}) locally, but it is obviously discontinuous across $x=0$. 

We can also find a large set of additional local solutions, which for the same fixed $E$ depend on an additional real parameter $k\geq 0$. Defining the auxiliary quantity
\begin{eqnarray}\label{kappa}
\kappa(k) = \sqrt{U^{2}-E^{2}+k^{2}}
\end{eqnarray}
we can express these local solutions as $\Psi\to\psi_{k}$ for\begin{widetext}
\begin{eqnarray}\label{psik}
\psi_{k} &=& e^{-\kappa(k)|x|}\left[\,\mathrm{sgn}(x) - \frac{iE}{\kappa(k)}\right]\left[\left(\begin{matrix} 1 \\ 1\end{matrix}\right)[\cos(ky)+\,\mathrm{sgn}(x)\frac{U}{k}\sin|ky|]-i\left(\begin{matrix} 1 \\ -1\end{matrix}\right) \frac{\,\mathrm{sgn}(x)\kappa(k) +iE}{k}\sin(ky)\right]\nonumber\\
&=& e^{-\kappa(k)|x|}\left[\left(\frac{U}{k}\sin(k|y|) -\frac{iE}{\kappa(k)}\cos(ky)\right)\left(\begin{matrix} 1 \\ 1\end{matrix}\right)-i\frac{U^{2}+k^{2}}{k\kappa(k)}\sin(ky)\left(\begin{matrix} 1 \\ -1\end{matrix}\right)\right]+\mathrm{sgn}(x)e^{-\kappa(k)|x|}\Delta\chi_{k}(|y|)\nonumber\\
\Delta\chi_{k}(|y|)&=&\left(\cos(ky)-\frac{iEU}{k\kappa(k)}\sin(k|y|)\right)\left(\begin{matrix} 1 \\ 1\end{matrix}\right)
 \;.
\end{eqnarray}
\end{widetext}
One can verify directly that these 2-spinors are all local solutions of (\ref{TDD}) with eigenvalue $E$, and they are manifestly continuous across $y=0$. The discontinuity across $x=0$, however, is $2\Delta\chi_{k}(|y|)\not=0$.

Since the discontinuities across $x=0$ of both $\Psi_{E-}$ and all the $\psi_{k}$ are even functions of $y$, however, and proportional to the same spinor $(1,1)^{T}$, we can construct our single continuous eigenfunction $\Psi_{E}$ as
\begin{eqnarray}\label{PsiE}
\Psi_{E}(x,y) = \Psi_{E-}(x,y) +\int_{0}^{\infty}\!dk\,f_{E}(k)\,\psi_{k}(x,y)
\end{eqnarray}
if only we can find a function $f_{E}(k)$ such that
\begin{eqnarray}\label{IntEq}
\int_{0}^{\infty}\!dk\,f_{E}(k)\left[\cos(ky) - i\frac{EU}{k\kappa(k)}\sin(k|y|)\right] =\frac{e^{-U|y|}}{2}\;.
\end{eqnarray}

Since both $\Psi\to\Psi_{E-}$ and $\Psi\to\psi_{k}$ satisfy $\Psi = \sigma_{x}\Psi(x,-y)\equiv \Psi_{R1} $, this $\Psi_{E}$ also satisfies $\Psi_{E}(x,y) = \sigma_{x}\Psi_{E}(x,-y)$. The orthogonal second eigenspinor with the same energy eigenvalue $E$ will then be given by the $R_{2}$ reflection, $\tilde{\Psi}_{E}=\sigma_{y}\Psi_{E}(-x,y)$. 

The checkerboard Dirac problem therefore reduces to solving the integral equation (\ref{IntEq}) for $f_{E}(k)$, with $0\leq E\leq U$ assumed without loss of generality.

\section{Solving the integral equation}
\subsection{Basic approach}
Physicists are so accustomed to solving differential equations that it can be difficult even to approach an integral equation such as (\ref{IntEq}). One is apt to keep thinking that the challenge is still the more usual one, of finding the unknown integral of a given integrand. One can therefore spend a long time constructing ingenious forms of $f_{E}(k)$ which have the merit that they allow the $k$-integration in (\ref{IntEq}) to be performed analytically for all $y$. Each time one has constructed such a conveniently integrable $f_{E}$, one feels that it is a discovery which must be a valuable advance toward solution. In fact, though, merely finding an $f_{E}$ which lets one evaluate the integral is only as big a step towards solving the integral equation as the step one makes toward solving a differential equation, when one finds a function which one knows how to differentiate. In other words, it is usually a worthless step, because to solve a differential equation, one must not merely find a function whose derivatives can be evaluated: one must find the function whose derivatives satisfy the equation. So it is for the integral equation (\ref{IntEq}): the challenge is not merely to find an $f_{E}(k)$ which lets us evaluate the integrals in the equation, but to find that very particular $f_{E}(k)$ whose integrals actually satisfy the equation --- for all values of $y$. 

Integral equations are thus generally harder than differential equations, because the task of finding functions to satisfy the equation is the same, but the straightforward step of differentiation has been replaced by the generally harder task of integration. One basic strategy which can successfully be applied to both kinds of equation, however, is the adaptation approach of modifying a solution which is known for a simple limit into a solution of the general equation. This strategy requires an initial limiting solution, and in our case, we can form one by considering the simple case $E=0$, for which the residue theorem of Cauchy supplies the obvious solution
\begin{eqnarray}\label{f0}
f_{0}(k) = \frac{U}{2\pi(k^{2}+U^{2})}\;.
\end{eqnarray}

This $f_{0}(k)$ solution works because it is an even function of $k$, allowing us to extend the lower limit of the $k$-integral in (\ref{IntEq}) from 0 to $-\infty$ and replace $\cos(ky)\to e^{ik|y|}$, and has a simple pole in the upper complex $k$-plane, at $k=iU$. Our basic approach will therefore be to construct a more general $f_{E}(k)$ which also lets us evaluate the integral in (\ref{IntEq}) by contour integration, and satisfies the equation with a residue at $k=iU$. The form of $f_{E}(k)$ which is required for $E\not= 0$ will turn out to be complicated, but we can anticipate many of its features by considering the large $|y|$ limit of (\ref{IntEq}).

\subsection{Deductions from large $|y|$}
The right-hand side of (\ref{IntEq}) vanishes exponentially at large $y$, and the method of stationary phase (or the theory of Fourier transforms) tells us that the integral on the left-hand side will be dominated for large $|y|$ by small $k$. We can therefore write
\begin{eqnarray}
	\lim_{|y|\to\infty} e^{-U|y|} = - 2i\frac{EU}{\sqrt{U^{2}-E^{2}}}\int_{0}^{\infty}\!dk\,\frac{f_{E}(k)}{k}\sin(k|y|)\;.
\end{eqnarray}
But then we can note that
\begin{eqnarray}
	\int_{0}^{\infty}\!dk\, \frac{\sin(k|y|)}{k} = \frac{\pi}{2}\;, 
\end{eqnarray}
which is not zero for large $y$. Hence in order to have this integral correctly vanish at large $y$, we must have $\lim_{k\to0}f_{E}(k) = 0$.

We can also note that for integer $n\geq 0$
\begin{eqnarray}
	\lim_{\epsilon\to0^{+}}\int_{0}^{\infty}\!dk\, e^{-\epsilon k}k^{2n+1}\sin(ky) = -(-1)^{n}\pi \frac{d^{2n+1}}{dy^{2n+1}}\delta(y) \nonumber
\end{eqnarray}
indeed vanishes at large $y$, so that even powers of $k$ are allowed in the Taylor expansion of $f_{E}(k)$ about $k=0$; but for odd powers of $k$ we instead have
\begin{eqnarray}
\lim_{\epsilon\to0^{+}}\int_{0}^{\infty}\!dk\, e^{-\epsilon k}k^{2n}\sin(ky) \sim y^{-n}	
\end{eqnarray}
which vanishes at large $y$, but too slowly to be of exponential form $e^{-U|y|}$. Hence there cannot be any odd powers of $k$ in $f_{E}(k)$: it must be an even function, not just for $E=0$ but also for $E>0$.

Since $f_{E}(-k) = f_{E}(k)$, then, we can note that the factor
\begin{equation}
\left[\cos(ky) - i\frac{EU}{k\kappa(k)}\sin (k|y|)\right]
\end{equation}
is itself an even function of $k$, and so we can re-write (\ref{IntEq}) equivalently as
\begin{eqnarray}\label{IntEq2}
	\int_{-\infty}^{\infty}\!dk\,f_{E}(k)\,e^{ik|y|}\left[1 - \frac{EU}{k\sqrt{U^{2}-E^{2}+k^{2}}}\right] = e^{-U|y|}\;.
\end{eqnarray}
We can therefore solve (\ref{IntEq2}) by finding an even $f_{E}(k)$ such that the product
\begin{eqnarray}\label{product}
	X_{E}(k) &=& f_{E}(k)\left(1-\frac{EU}{k\sqrt{U^{2}-E^{2}+k^{2}}}\right)
\end{eqnarray}
 is analytic everywhere in the upper half of the complex $k$-plane, except for a pole at $k=iU$. 
  
Unfortunately, we cannot just do this in the obvious way, by merely setting
 \begin{eqnarray}\label{try}
	f_{E}(k) \overset{?}{\longrightarrow} \frac{U}{2\pi(k^{2}+U^{2})} \left(1-\frac{EU}{k\sqrt{U^{2}-E^{2}+k^{2}}}\right)^{-1}\;,\nonumber
\end{eqnarray}
because although this would solve (\ref{IntEq2}), it would not solve the equation (\ref{IntEq}) which we really need to solve, because the resulting $f_{E}(k)$ would not be an even function of $k$ (except for $E=0$). Hence (\ref{IntEq2}) would not be equivalent to (\ref{IntEq}) for this $f_{E}(k)$. In fact, as we have seen, only an even function of $k$ can solve (\ref{IntEq}).

We have, however, learned enough about the behavior of $f_{E}(k)$ that we will now be able to determine it for general $E$. Our main concern will be the branch point in $\sqrt{U^{2}-E^{2}+k^{2}}$ at $k=i\sqrt{U^{2}-E^{2}}$. This requires a branch cut in the complex $k$-plane. To keep the branch cut away from the real $k$-axis along which we must actually integrate, and to place it symmetrically, we will let it run vertically up the imaginary $k$-axis from the branch point. We will solve (\ref{IntEq2}) using the residue theorem, just as we did for $E=0$, but in order to apply it we will first have to remove the branch cut in $X_{E}(k)$, by funding an even function $f_{E}(k)$ which has an exactly complementary branch cut discontinuity to that of the other factor in $X_{E}(k)$.

\subsection{General form of $f_{E}(k)$}

Our general $f_{E}(k)$ must be of this form:
\begin{widetext}\begin{eqnarray}\label{fEform}
f_{E}(k) =  Z(E)\frac{\exp\left[ -i\phi_{E}(\sinh^{-1}\left(\frac{k}{\sqrt{U^{2}-E^{2}}}\right)\right]}{\pi(k^{2}+U^{2})}\frac{k^{2}}{k^{2}-(E+i0)^{2}}\frac{\sqrt{U^{2}-E^{2}+k^{2}}}{\sqrt{k^{2}+U^{2}}}
\end{eqnarray}
\end{widetext}
where $E+i0$ is short for $\lim_{\epsilon\to0^{+}}(E+i\epsilon)$, $\phi_{E}(\sigma)$ is a certain non-trivial function that will be defined below, and $Z(E)$ is a normalization factor which ensures that the residue which survives the integration in (\ref{IntEq}) is exactly $i \pi e^{-U|y|}$. Why does $f_{E}(k)$ take this form?

First of all, it is an even function of $k$ ($\phi_{E}$ is an even function of its argument). Secondly, it vanishes quadratically as $k\to0$. Thirdly, it also vanishes with $k^{-2}$ as $k\to\infty$, so that we will be able to close contours at infinity in the complex $k$-plane. Fourthly, it has a pole at $k=iU$, and fifthly, its only other pole in the upper half of the complex $k$-plane is at $k=E+i0$ --- but at this pole the factor $1- EU/[k\kappa(k)]$ in the (\ref{IntEq2}) integrand has a zero, so that the whole integrand $X_{E}(k)$ is regular at $k=E$, and therefore no residue $\sim e^{iE|y|}$ will appear and ruin our solution to (\ref{IntEq}).

These five features explain all the factors in (\ref{fEform}), except for the final factor of $\sqrt{U^{2}-E^{2}+k^{2}}/\sqrt{U^{2}+k^{2}}$. This final factor is simply cosmetic, since it could have been absorbed into the non-trivial function $\phi_{E}$; with the benefit of hindsight, knowing how complicated $\phi_{E}$ turns out to be, we have chosen to simplify $\phi_{E}$ by extracting this factor. The $\sqrt{U^{2}-E^{2}+k^{2}}$ should at least make sense as a potentially useful factor in $f_{E}(k)$, since it already appears in the $X_{E}(k)$ integrand (\ref{product}). The $\sqrt{k^{2}+U^{2}}$ in the denominator is then inserted in order to cancel the $\sqrt{U^{2}-E^{2}+k^{2}}$ for $E\to0$, because this then lets us recover the simple known $f_{0}(k)$ with the simple limit $\phi_{0}(k)=0$.

Because our analysis will mainly concern the branch cut, it will simplify many expressions if we extract the simple poles from $X_{E}(k)$ by defining\begin{widetext}
\begin{eqnarray}\label{BEkdef}
X_{E}(k) &=& \frac{Z(E) k \tilde{X}_{E}(k)} {\pi(k^{2}+U^{2})[k^{2}-(E+i0)^{2}]} \nonumber\\
\tilde{X}_{E}(k) &=& \exp\left[ -i\phi_{E}(\sinh^{-1}\left(\frac{k}{\sqrt{U^{2}-E^{2}}}\right)\right]\ \frac{EU-k\sqrt{U^{2}-E^{2}+k^{2}}}{\sqrt{k^{2}+U^{2}}}\equiv e^{-i\phi_{E}}B_{E}(k)\;.
\end{eqnarray}\end{widetext}
Our task has therefore become that of constructing an even function $\phi_{E}$ such that the product $\tilde{X}_{E}(k)=e^{-i\phi_{E}}B_{E}(k)$ is analytic in the upper half of the complex $k$-plane.
Our whole problem now turns on the as-yet-unspecified factor $e^{-i\phi_{E}}$, whose argument we have defined to be
\begin{eqnarray}\label{sigmadef}
\sigma(k)=\sinh^{-1}\left(\frac{k}{\sqrt{U^{2}-E^{2}}}\right)\;.
\end{eqnarray}
Since $\phi_{E}$ is still at this point an arbitrary function, why have we specified that it depends on $k$ through this particular $\sigma(k)$? The reason is that the whole difficulty in solving the integral equation (\ref{IntEq2}) concerns the branch cut in $B_{E}(k)$, which runs upwards from $k=i\sqrt{U^{2}-E^{2}}$. Defining the argument $\sigma(k)$ as we have will enable us to deal with this branch cut more easily, by mapping it simply.

\begin{figure}[H]
\includegraphics[width=.45\textwidth]{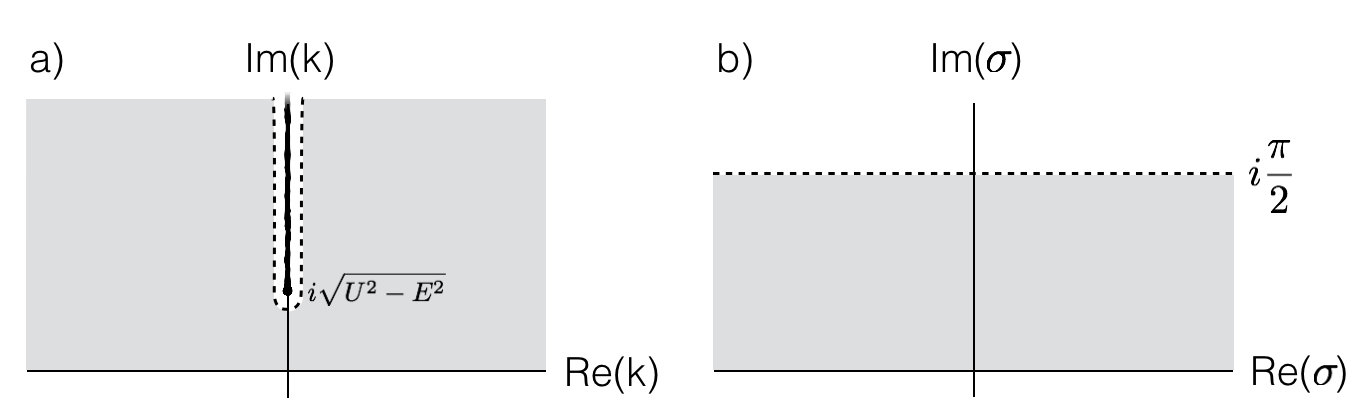}
\caption{The co-ordinate mapping $\sigma(k)$ of the upper half of the complex $k$-plane (a) into the strip $0\leq\mathrm{Im}(\sigma)\leq \pi/2$ in the complex $\sigma$-plane (b). The thick rough line in (a) is the branch cut. The two sides of branch cut in (a) are `unfolded' into the full dashed line $\mathrm{Im}(\sigma)=\pi/2$ in (b). Functions of $k$ which are continuous across the branch cut are thereby represented as functions of complex $\sigma$ which, when considered as functions on the horizontal line $\mathrm{Im}(\sigma)=\pi/2$, are even functions of the argument $\mathrm{Re}(\sigma)$.}
\end{figure}

Because $\sinh(k+i\pi)\equiv - \sinh(k)$ for all complex $k$, the entire complex $k$-plane is mapped into the strip $|\mathrm{Im}(\sigma)|\leq\pi/2$ of complex $\sigma$. The upper half of the complex $k$-plane is mapped into the upper half of this $\sigma$ strip, and what is especially important is that the horizontal line $\sigma = s + i\pi/2$, for real $s$, represents \textit{both sides}Ê of the branch cut. To see this, we note that the contour $k\to k_{B}(s)$, with
\begin{eqnarray}\label{branchcontour}
k_{B}(s) &=& \lim_{\epsilon\to0^{+}}\sqrt{U^{2}-E^{2}}[i\cosh(s)+\epsilon \sinh(s)]\nonumber\\
&=& \lim_{\epsilon\to0^{+}}\sqrt{U^{2}-E^{2}}\sinh\left(s+i\left(\frac{\pi}{2}-\epsilon\right)\right)
\end{eqnarray}
for real $s\in(-\infty,\infty)$, runs down the left side of the branch cut for $s<0$, and then back up the right side of the branch cut for $s>0$. Hence every two points $\sigma = \pm s + i\pi/2$ for real $s$ correspond to two complex values of $k$ directly across the branch cut from each other. 

It is this latter fact which provides the point of the $\sigma$ mapping: functions with discontinuities across the branch cut in the complex $k$-plane correspond to functions in the complex $\sigma$-strip which are not even functions of real $s$ along the line $\sigma=s+i\pi/2$. In particular our integrand factor $B_{E}(k)$ along the branch cut contour $k\to k_{B}(s)$ is given by
\begin{eqnarray}\label{BEk}
B_{E}(k_{B}(s))&=& \frac{EU-k_{B}(s)\sqrt{U^{2}-E^{2}+k_{B}^{2}(s)}}{\sqrt{k_{B}^{2}(s)+U^{2}}}\\
&=&-\cosh(s) \sqrt{U^{2}-E^{2}\tanh^{2}(s)}e^{-i\beta(s)} e^{D(s)}\nonumber
\end{eqnarray}
for
\begin{eqnarray}
\beta(s) &=&\frac{\pi}{2} \theta(|s|-\tanh^{-1}(E/U)) \\
D(s) &=& \frac{1}{2}\ln\left(\frac{|E+U\tanh(s)|\,|U-E\tanh(s)|}{|E-U\tanh(s)|\,|U+E\tanh(s)|}\right)\;.\nonumber
\end{eqnarray}
Here $\theta(x)$ is the usual step function, equalling one for positive arguments and zero for negative. The only factor in $B_{E}(k_{B}(s))$ which is not even in $s$ is $e^{D(s)}$; $D(s)=-D(-s)$ is manifestly an \textit{odd} function of $s$. 

The fact that the branch discontinuity of $B_{E}(k)$ can thus be expressed entirely in the oddness of the exponent $D(s)$ will be the basis for our construction of $\phi_{E}(\sigma)$. While keeping $\phi_{E}$ an \textit{even} function of complex $\sigma$, we will construct it to have a part which is \textit{odd} in the \textit{real part} of $\sigma$ along the line $\mathrm{Im}(\sigma)=\pi/2$, in just such a way as to repair the branch cut in the combined integrand $X_{E}(k)$ in Eqn.~(\ref{IntEq2}), and thereby solve the integral equation to obtain our continuous Dirac eigenspinors. Given the evenness of $\phi_{E}$ in the whole $\sigma$-plane, specifying its odd part along the line $\mathrm{Im}(\sigma)=\pi/2$ will uniquely determine the whole function.

(The discontinuities in $B_{E}(k_{B}(s))$ at the points $s = \pm\tanh^{-1}(E/U)$, which correspond to $k= iU\pm0$, are also non-analytic features which must be repaired in the product $\tilde{X}_{E}(k)$, even though these discontinuous steps occur along the branch cut rather than across it; but it will turn out conveniently that these discontinuities will also be repaired by the $\phi_{E}(\sigma)$ that we will obtain when we repair the discontinuity across the branch cut that is expressed in $D(s)$.)\\

\subsection{Repairing the branch cut}
We have now expressed the as-yet-undetermined factor in $f_{E}(k)$, as well as the branch cut discontinuity in $B_{E}(k_{B}(s))$, through the respective exponents $i\phi_{E}(\sigma(k))$, and $D(s)$. This means that we now only need to find a $i\phi_{E}$ whose odd part along $\sigma = s + i\pi/2$ cancels $D(s)$. The total of the two exponents will then be an even function of $s$ along this line, and therefore there will be no net branch discontinuity in the product $\tilde{X}_{E}(k)=e^{-i\phi_{E}}B_{E}(k)$. With the branch cut thus gone, $X_{E}(k)$ will be analytic in the upper half $k$-plane except for the simple pole at $k=iU$, and residue calculus will confirm that the integral equation is satisfied, once the normalization constant $Z(E)$ in (\ref{fEform}) has been chosen appropriately. In this sense, we are now in a position to solve the integral equation (\ref{IntEq}) by a simple subtraction. This turns out to really be almost as straightforward as it sounds, because of two seemingly complicated Fourier integrals that can both be computed explicitly.

The first integral is the Fourier sine transform of the odd function $D(s)$, as defined in (\ref{BEk}) above:
\begin{eqnarray}
		D(s) &=& \int_{0}^{\infty}\!d\xi\, \Delta(\xi)\sin(s\xi)\nonumber\\
		\Delta(\xi)&=& -\frac{1}{\pi}\int_{0}^{\infty}\!ds\,\sin(\xi s)\, D(s)\nonumber\\
&=& \frac{\tanh\left(\frac{\pi}{4}\xi\right)\sin\left(\xi\tanh^{-1}(E/U)\right)}{\xi}\;.
\end{eqnarray}
This integral can be obtained by routine methods, first integrating by parts and then computing the resulting integrals of rational functions of hyperbolic functions using standard residue techniques. Contour methods likewise confirm the inverse Fourier integral.

We then note that since $\phi_{E}(\sigma)$ must be an even function of $\sigma$, it makes sense to define $\phi_{E}(\sigma)$ by defining its Fourier cosine transform $F(\xi)$, such that 
\begin{eqnarray}\label{Fs}
\phi_{E}(\sigma)=\int_{0}^{\infty}\!d\xi\,F(\xi)\cos(\sigma\xi)\;.
\end{eqnarray}
Exactly what function $F(\xi)$ is, is still up to us to construct; but just from the form of the cosine transform we can note that along the branch cut contour $\sigma(k_{B}(s))= s+i\pi/2$ we will have\begin{widetext}
\begin{eqnarray}\label{branchphi}
		\phi_{E}\left(i\frac{\pi}{2} +s\right) &=& \int_{0}^{\infty}\!d\xi\, F(\xi)\cos\left(\xi\left[i\frac{\pi}{2} +s\right]\right)
		=\int_{0}^{\infty}\!d\xi\, F(\xi)\left[\cosh\left(\frac{\pi}{2}\xi\right)\cos(s\xi) - i \sinh\left(\frac{\pi}{2}\xi\right)\sin(s\xi)\right] \;.
\end{eqnarray}

The part of $\phi_{E}(s+i\pi/2)$ which is odd in $s$ is clearly given by the sine part of the final integral in (\ref{branchphi}). We can therefore make the odd part of $i\phi_{E}(s+i\pi/2)$ exactly cancel $D(s)$ simply by choosing
\begin{eqnarray}\label{solution1}
F(\xi) = -\frac{\Delta(\xi)}{\sinh\left(\frac{\pi}{2}\xi\right)} = \frac{\sin\left(\xi\tanh^{-1}(E/U)\right)}{2\xi\cosh^{2}\left(\frac{\pi}{4}\xi\right)}\;.
\end{eqnarray}
With this prescription, the product $\exp(i\phi_{E}(i\pi/2 + s)\,\exp(D(s))$ will be purely even in $s$, and hence the total integrand $X_{E}(k)=f_{E}(k)B_{E}(k)$ in (\ref{IntEq2}) will be continuous across the branch cut that is present in $B_{E}(k)$ alone. 

Amazingly enough, we can then perform the Fourier cosine transform (\ref{Fs}) with this $F(\xi)$ and obtain $\phi_{E}(\sigma)$ explicitly, albeit in terms of a moderately obscure special function.

\subsection{Explicit solution for $\phi_{E}(\sigma)$}
The integral (\ref{Fs}) can be performed by first differentiating under the integral with respect to $\tanh^{-1}(E/U)$, to obtain $\partial \phi_{E}/\partial(\tanh^{-1}(E/U)$ in terms of some Fourier transforms of $1/\cosh^{2}(\pi \xi/4)$; these can be obtained by contour methods or simply looked up in tables. Integrating then with respect to $\tanh^{-1}(E/U)$, to obtain $\phi_{E}(\sigma)$ from its partial derivative, requires indefinite integrals of $x/\sinh(x)$. After some integration by parts, these lead to functions known as \textit{dilogarithms}; the final result for $\phi_{E}(\sigma)$ is
\begin{eqnarray}\label{phisol}
\phi_{E}(\sigma) &=& \frac{1}{2\pi}\left[Q\left(2(\sigma +\tanh^{-1}\frac{E}{U})\right)-Q\left(2(\sigma -\tanh^{-1}\frac{E}{U})\right)\right]\\
Q(x) &:=& x\ln\frac{1-e^{-x}}{1+e^{-x}} - \,\mathrm{Li}_{2}(e^{-x}) + \,\mathrm{Li}_{2}(-e^{-x})\nonumber\\
\,\mathrm{Li}_{2}(z) &:=& -\int_{0}^{z}\!du\,\frac{\ln(1-u)}{u}= \sum_{n=1}^{\infty}\frac{z^{n}}{n^{2}} \;\;\hbox{where this converges}\nonumber\;.
\end{eqnarray}
The function $\,\mathrm{Li}_{2}(z)$ is the dilogarithm \cite{Maximon2003}.

The dilogarithm obeys many convenient identities \cite{Maximon2003}, including the \textit{inversion identity} by which one can prove that $\phi_{E}(\sigma)$ is indeed an even function of $\sigma$, even though this is not immediately obvious in the expressions we have given:
\begin{eqnarray}\label{inversion}
\,\mathrm{Li}_{2}(z)+\,\mathrm{Li}_{2}(1/z)=-\frac{\pi^{2}}{6}-\frac{1}{2}[\ln(-z)]^{2}\;.
\end{eqnarray}
The dilogarithm $\,\mathrm{Li}_{2}(z)$ has a logarithm-like branch point at $z=1$, and is otherwise analytic; the usual convention is to run the branch cut to the right along the real $z$ axis \cite{Maximon2003}. From the inversion identity and the fact that $\,\mathrm{Li}_{2}(z)$ is analytic around the real axis with $\mathrm{Re}(z)<1$, one can derive the branch discontinuity; if we write $z = e^{x\pm i\epsilon}$ then the branch cut runs along real $x>0$ and from (\ref{inversion}) we obtain
\begin{eqnarray}\label{inversion2}
\,\mathrm{Li}_{2}(e^{(x\pm i\epsilon)})&=&-\,\mathrm{Li}_{2}(e^{-x})-\frac{\pi^{2}}{6}-\frac{1}{2}[\ln(e^{\mp i \pi}e^{(x\pm i\epsilon)})]^{2}\nonumber\\
&=&-\,\mathrm{Li}_{2}(e^{-x})-\frac{\pi^{2}}{6} -\frac{1}{2}[x\mp i\pi]^{2} \;.
\end{eqnarray}

This particular form of the branch discontinuity in the dilogarithm is exactly such that the combination $Q(z)$ is analytic for all 
$|\mathrm{Im}(z)|<(\pi/2)$. Hence $f_{E}(k)$ as given by (\ref{fEform}) is indeed analytic everywhere except along the $B_{E}(k)$ branch cut (and at the pole $k=iU$). In particular, $\phi_{E}(\sigma)$ is real, even, and well-behaved along the entire real axis of $\sigma$, which coincides with the real axis of $k$. We can now confirm explicitly that our rather complicated $\phi_{E}(\sigma)$ really does remove the $B_{E}(k)$ branch cut, and as an added bonus, also removes the branch point at $k=iU$ in $f_{E}(k)$, leaving only our desired simple pole.

Approaching the $B_{E}(k)$ branch cut, which means $\sigma = i\pi/2 + s$ with $s$ running infinitesimally below the real axis, we can see from its definition that
\begin{eqnarray}
\phi_{E}\left(i\frac{\pi}{2}+s\right) &=& -\phi_{E}(s) +\frac{i}{2}\ln\left(\frac{\left[1+e^{-2[s +\tanh^{-1}(E/U)]}\right]\,\left[1-e^{-2[s -\tanh^{-1}(E/U)]}\right]}{\left[1+e^{-2[s -\tanh^{-1}(E/U)]}\right]\,\left[1-e^{-2[s +\tanh^{-1}(E/U)]}\right]}\right)\nonumber\\
&=& -\phi_{E}(s) +\frac{i}{2}\ln\left(\frac{\left[1+e^{-2[s +\tanh^{-1}(E/U)]}\right]\,\left|1-e^{-2[s -\tanh^{-1}(E/U)]}\right|}{\left[1+e^{-2[s -\tanh^{-1}(E/U)]}\right]\,\left|1-e^{-2[s+\tanh^{-1}(E/U)]}\right|}\right) \nonumber\\
&&+\frac{\pi}{2}[\theta(\tanh^{-1}(E/U)-s) - \theta(-\tanh^{-1}(E/U)-s)]\nonumber\\
&\equiv&  -\phi_{E}(s) + \frac{\pi}{2}-\beta(s)+iD(s)\;.
\end{eqnarray}
Using (\ref{BEk}), we can then confirm that along the branch cut the pole-free factor $\tilde{X}_{E}(k)$ in the full integrand $X_{E}(k)$ is
\begin{eqnarray}\label{checkit}
\tilde{X}_{E}(k_{B}(s)) &=& e^{-i\phi_{E}(s+i\pi/2)}B_{E}(k) = -\cosh(s) \sqrt{U^{2}-E^{2}\tanh^{2}(s)}e^{i[\phi_{E}(s+i\pi/2)-\beta(s)-iD(s)]}\nonumber\\
&\equiv& -i\cosh(s) \sqrt{U^{2}-E^{2}\tanh^{2}(s)} e^{-i\phi_{E}(s)}
\end{eqnarray}
which is an even and regular function of $s$. The pole-free part $\tilde{X}_{E}(k)$ of the (\ref{IntEq2}) integrand $X_{E}(k)$ is therefore analytic throughout the upper half of the complex $k$-plane, and the integral in (\ref{IntEq2}) can be computed by closing the integration contour at upper infinity and summing residues at the poles in $X_{E}(k)$. Since $\tilde{X}_{E}(E)=0$, there is no residue at the pole $k=E+i0$; the entire integral is given by the residue at $k=iU$. This provides the correct $e^{-U|y|}$ dependence on $y$ to solve the integral equation; all that remains is to fix the normalization constant $Z(E)$.

\subsection{Normalization}
The point $k=iU$ corresponds to either of $\sigma = i\pi/2 \pm \tanh^{-1}(E/U)-i\epsilon$: our elimination of the branch cut guarantees that the residue is identical at both these points in $\sigma$. If for example we choose the $+$ option, approaching the $k=iU$ from the upper right, then using $\tanh^{-1}(E/U) = \ln\sqrt{(U+E)/(U-E)}$ we obtain the condition
\begin{eqnarray}
\frac{1}{2\pi i}&=& \frac{Z(E)}{\pi}\frac{EU}{U^{2}+E^{2}}e^{+i\phi_{E}\left(\tanh^{-1}(E/U)\right)}\times\lim_{\delta \to 0^{+}}\left[\sqrt{\frac{1+\left(\frac{U-E}{U+E}\right)^{2}}{1-\left(\frac{U-E}{U+E}\right)^{2}}}\frac{\sqrt{1-e^{-2\delta}}}{2i\sqrt{UE}\sqrt{\delta}}\right]\nonumber\\
&=& \frac{Z(E)}{2\pi i}\frac{e^{+i\phi_{E}\left(\tanh^{-1}(E/U)\right)}}{\sqrt{U^{2}+E^{2}}}
\end{eqnarray}
 \end{widetext}
This yields
\begin{eqnarray}\label{norm}
Z(E) &=& e^{-i\phi_{E}\left(\tanh^{-1}(E/U)\right)}\sqrt{U^{2}+E^{2}}\;.
\end{eqnarray}

This completes our integral representation of the Dirac zero-line mode eigenspinor $\Psi_{E}(x,y)$, which solves the time-independent Dirac equation (\ref{Dirac1}) with the checkerboard potential.
The eigenspinor is given by the integral representation (\ref{PsiE}), in terms of the local spinors $\psi_{k}(x,y)$ given by (\ref{psik}) and the weighting function $f_{E}(k)$ as defined by (\ref{fEform}), with the nontrivial phase function $\phi_{E}$ defined by (\ref{phisol}). 

\section{Summary and Plots}
\subsection{Summary of results}
Because our result combines sub-results in several equations throughout our text, we collect them all here. We have solved the two-dimensional time-independent Dirac equation with a checkerboard potential, 
\begin{eqnarray}
E \Psi_{E}(x,y) = U\mathrm{sgn}(x)\mathrm{sgn}(y)\sigma_{z}\Psi_{E} -i\left[ \sigma_{x}\partial_{x}+\sigma_{y}\partial_{y}\right]\Psi_{E}\;.\nonumber
\end{eqnarray}
We can also solve the different Dirac equation that is obtained by reversing the sign of the $\partial_{x}$ term, by taking $\Psi_{E}\to\Psi^{*}_{E}$. This other equation would describe graphene modes in the neighborhood of the other Dirac point.

We have assumed $U>0$ without loss of generality, because for negative $U$ the transformation $\Psi_{E}(x,y)\to\sigma_{z}\Psi_{E}$ and $E\to -E$ restores positive $U$. We have assumed $|E|<U$ because we are only interested here in zero-line modes, which have energies within the bulk energy gap and which are bound to the $V=0$ channels along the $x$ and $y$ axes. We have further assumed $E\geq 0$ without loss generality, because once all the positive energy $\Psi_{E}$ have been found, the negative energy solutions are obtained as $\Psi_{-E}= \sigma_{x}\Psi_{E}^{*}$. Finally, we have chosen only one of the two degenerate eigenspinors that exist for each $E$, namely the one which satisfies $\sigma_{x}\Psi_{E}(x,-y)=\Psi_{E}(x,y)$. The orthogonal second solution is given by $\tilde{\Psi}_{E}(x,y)=\sigma_{y}\Psi_{E}(-x,y)$, and it satisfies $\sigma_{x}\tilde{\Psi}_{E}(x,-y)=-\tilde{\Psi}_{E}(x,y)$.

Our solution $\Psi_{E}$ is built up from discontinuous local solutions, namely
\begin{eqnarray}\Psi_{E-} = \left\{
\begin{matrix}e^{iE x}e^{-U|y|}\left(\begin{matrix} 1\\ 1\end{matrix}\right)&,\; x<0\\ 0 &,\; x>0 \end{matrix}\right.
\end{eqnarray}
and
\begin{widetext}
\begin{eqnarray}\label{psikbis}
\psi_{k} &=& e^{-\kappa(k)|x|}\left[\,\mathrm{sgn}(x) - \frac{iE}{\kappa(k)}\right]\left[\left(\begin{matrix} 1 \\ 1\end{matrix}\right)[\cos(ky)+\,\mathrm{sgn}(x)\frac{U}{k}\sin|ky|]-i\left(\begin{matrix} 1 \\ -1\end{matrix}\right) \frac{\,\mathrm{sgn}(x)\kappa(k) +iE}{k}\sin(ky)\right]\end{eqnarray}
where $\kappa(k) = \sqrt{U^{2}-E^{2}+k^{2}}$. In terms of these local solutions, the continuous global solution is defined as
\begin{eqnarray}\label{PsiEbis}
\Psi_{E}(x,y) = \Psi_{E-}(x,y) +\int_{0}^{\infty}\!dk\,f_{E}(k)\,\psi_{k}(x,y)
\end{eqnarray}
for the weight function
\begin{eqnarray}\label{fEformbis}
f_{E}(k) =  \sqrt{U^{2}+E^{2}}\, e^{-i\phi_{E}\left(\tanh^{-1}(E/U)\right)}\frac{\exp\left[ -i\phi_{E}(\sinh^{-1}\left(\frac{k}{\sqrt{U^{2}-E^{2}}}\right)\right]}{\pi(k^{2}+U^{2})}\frac{k^{2}}{k^{2}-(E+i0)^{2}}\frac{\sqrt{U^{2}-E^{2}+k^{2}}}{\sqrt{k^{2}+U^{2}}}\;.
\end{eqnarray}
Here the crucial phase function $\phi_{E}$ is defined as
\begin{eqnarray}\label{phisolbis}
\phi_{E}(\sigma) &=& \frac{1}{2\pi}\left[Q\left(2(\sigma +\tanh^{-1}\frac{E}{U})\right)-Q\left(2(\sigma -\tanh^{-1}\frac{E}{U})\right)\right]\\
Q(x) &=& x\ln\frac{1-e^{-x}}{1+e^{-x}} - \,\mathrm{Li}_{2}(e^{-x}) + \,\mathrm{Li}_{2}(-e^{-x})
\end{eqnarray}
where $\mathrm{Li}_{2}(z)$ is the dilogarithm function. It must be admitted that our result is somewhat formidable as an exact expression, but it is not really as complex as it is appears.

\subsection{Understanding $\Psi_{E}(x,y)$}
Because both $f_{E}$ and $\psi_{k}$ are even functions of $k$, we can extend the integral in (\ref{PsiEbis}) to minus infinity, and replace the sines and cosines with one exponential:
\begin{eqnarray}\label{tildepsik}
\int_{0}^{\infty}\!dk\,f_{E}(k)\psi_{k} &=& \frac{1}{2}\int_{-\infty}^{\infty}\!dk\,f_{E}(k)e^{-\kappa(k)|x|}e^{ik|y|}\left[\left(\begin{matrix} 1 \\ 1\end{matrix}\right)\left( \,\mathrm{sgn}(x) - \frac{iE}{\kappa(k)}\right)\left(1-i \,\mathrm{sgn}(x) \frac{U}{k} \right) -\,\mathrm{sgn}(y) \left(\begin{matrix} 1 \\ -1\end{matrix}\right)\frac{k^{2}+U^{2}}{k\kappa(k)}\right] \nonumber\\
&=&  -\frac{i}{2}\int_{-\infty}^{\infty}\!dk\,\frac{f_{E}(k)e^{-\kappa(k)|x|}e^{ik|y|}}{k\kappa(k)}\left[[Ek+U\kappa(k)]\left(\begin{matrix} 1 \\ 1\end{matrix}\right) + \,\mathrm{sgn}(y)(k^{2}+U^{2})\left(\begin{matrix} -i \\ i\end{matrix}\right)\right]\nonumber\\
&& +\frac{1}{2} \,\mathrm{sgn}(x)\left(\begin{matrix} 1 \\ 1\end{matrix}\right)\int_{-\infty}^{\infty}\!dk\,f_{E}(k)e^{-\kappa(k)|x|}e^{ik|y|}\left[1-\frac{EU}{k\kappa(k)}\right]\equiv \frac{1}{2}\int_{-\infty}^{\infty}\!dk\,f_{E}(k)\tilde{\psi}_{k}(x,y)\;.
\end{eqnarray}
If we then define $\eta = \tanh^{-1}(E/U)$, introduce polar coordinates $|x|=r\cos\theta$, $|y|=r\sin\theta$ for $0\leq\theta\leq\pi/2$, and perform the change of integration variable $k\to U\mathrm{sech}\,\eta\,\sinh\sigma$, these integrands are all analytic as functions of $\sigma$ for $0\leq \mathrm{Im}(\sigma)\leq \pi/2$ except for simple poles at $\sigma = \pm(\eta+i0)$ and $\sigma=\pm\eta + i\pi/2$.

It is convenient to integrate along the horizontal lines $\sigma = s + i\theta$ for real $s$, because then the position-dependent exponential $\exp(i k|y|-\kappa |x|) = \exp[- \mathrm{sech}(\eta) r \cosh(\sigma - i\theta)]$ becomes purely decaying instead of rapidly oscillatory. Moving the integration contour up from the real $\sigma$-axis to $\mathrm{Re}(\sigma)=i\theta$ crosses the pole at $\sigma = \eta+i0$; correcting for this requires adding the residue at the pole, which gives the asymptotic outoging zero-line modes along the $y$-axis. This leaves the general form that we promised back in (\ref{psiE}),
\begin{eqnarray}\label{psiE2}
\Psi_{E}(x,y) &=& \theta(-x) e^{i[Ex+\alpha(E/U)]}e^{-U|y|}\left(\begin{matrix}1\\ 1\end{matrix}\right)
+\frac{1}{2}e^{-U|x|}e^{i[E|y|-\alpha(E/U)]}\left[\left(\begin{matrix}1\\ 1\end{matrix}\right)-i\,\mathrm{sgn}(y)\left(\begin{matrix}1\\ -1\end{matrix}\right)\right]+\Delta\Psi_{E}(x,y)
\end{eqnarray}\end{widetext}
for a $\Delta\Psi_{E}$ whose dependence on $x^{2}+y^{2}$ comes entirely from the $\exp[- \mathrm{sech}(\eta) r \cosh(s)]$ factor in the integrands along lines of real $s$. For small $E/U=\tanh\eta$, therefore, $\Delta\Psi_{E}$ is localized within $r\lesssim 1/U$. As $E\to U^{-}$ this $r$-dependence disappears, and the zero-line modes spread out into the bulk 2D continuum; but since for large $\eta\sim-\ln\sqrt{1-E/U}$ this factor becomes $\exp[-r e^{s-\eta}]$, the spreading of $\Delta\Psi_{E}$ sets in only logarithmically as $E\to U$; it remains quite localized until $E$ is very close indeed to $U$.

(For $\theta=0$ and $\theta = \pi/2$, the convenient $\sigma = s + i\theta$ contours pass through the poles, but we can define the integrals by taking the principal value and then adding half-residues to correct for not avoiding the poles. The results are equivalent to taking the limits $\theta\to0^{+}$ or $\theta\to (\pi/2)^{-}$, and have the advantage that the principal value integration can be implemented numerically.)

The phase shift that we included in (\ref{psiE}) is now seen in (\ref{tildepsik}) to be given by $\alpha(E/U)=\phi_{E}\left(\tanh^{-1}(E/U)\right)$. This defines $\alpha(E/U)$ as rising smoothly from 0 at $E=0$ to $\pi/4$ at $E=1$. Its Taylor series in $E/U$ begins
\begin{eqnarray}
\alpha(x) = \frac{4x}{\pi} - \frac{20x^{3}}{9\pi} + \mathcal{O}(x^{5})\;;
\end{eqnarray}
since $\alpha$ is thus quite linear for small arguments, there is little dispersion associated with the intersection at low energies. Since the phase shift is negative, an incident packet emerges as a coherent pair of packets that are slightly \textit{ahead} of where they would be, if they travelled around the intersection's right angle corner at constant speed. The jump ahead is $(4/\pi)(1/U)$ at long wavelengths. The wave packets evidently cut the corner slightly.
\begin{figure}[H]
\includegraphics[width=.4\textwidth]{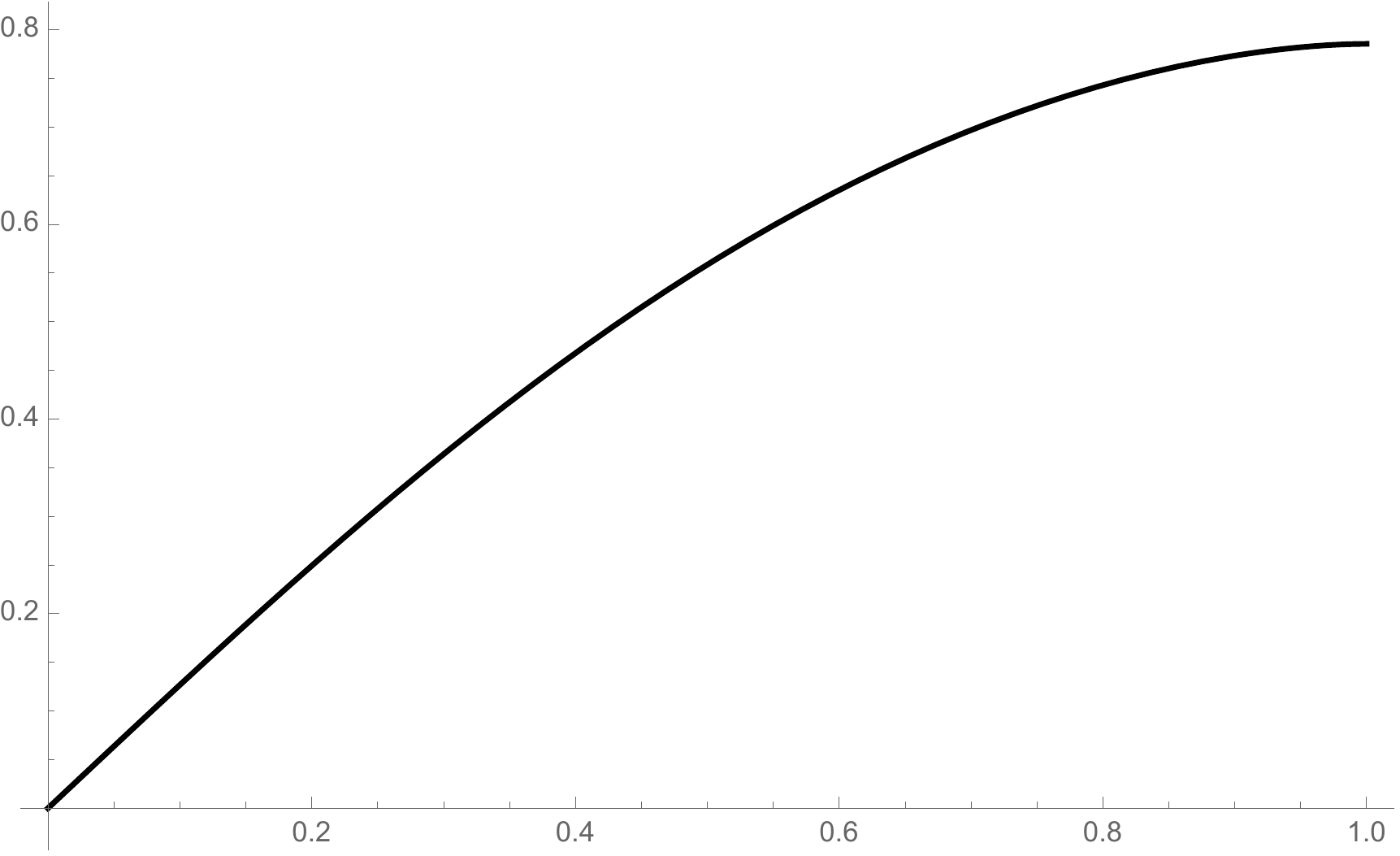}
\caption{The phase shift $\alpha(E/U)=2\phi_{E}(\tanh^{-1}(E/U)$.}
\end{figure}

\subsection{Plots}
Despite the complexity of our exact integral representations, they can be evaluated numerically without difficulty, and the spinor-valued functions which result turn out to be remarkably simple. The total density $|\Psi_{E}|^{2}$ is plotted in Fig.~4 as a function of $x$ and $y$ in units of $1/U$, for the cases $E/U = 0.1$, $0.5$, and $0.95$. The dependence on $E/U$ is remarkably slight, to the point where it is probably sufficient for most purposes to approximate $\Delta\Psi_{E}\doteq\Delta\Psi_{0}$ even for $E$ quite close to $U$. 
\begin{figure}[H]\label{FinFig}
\center\includegraphics[width=0.9\columnwidth]{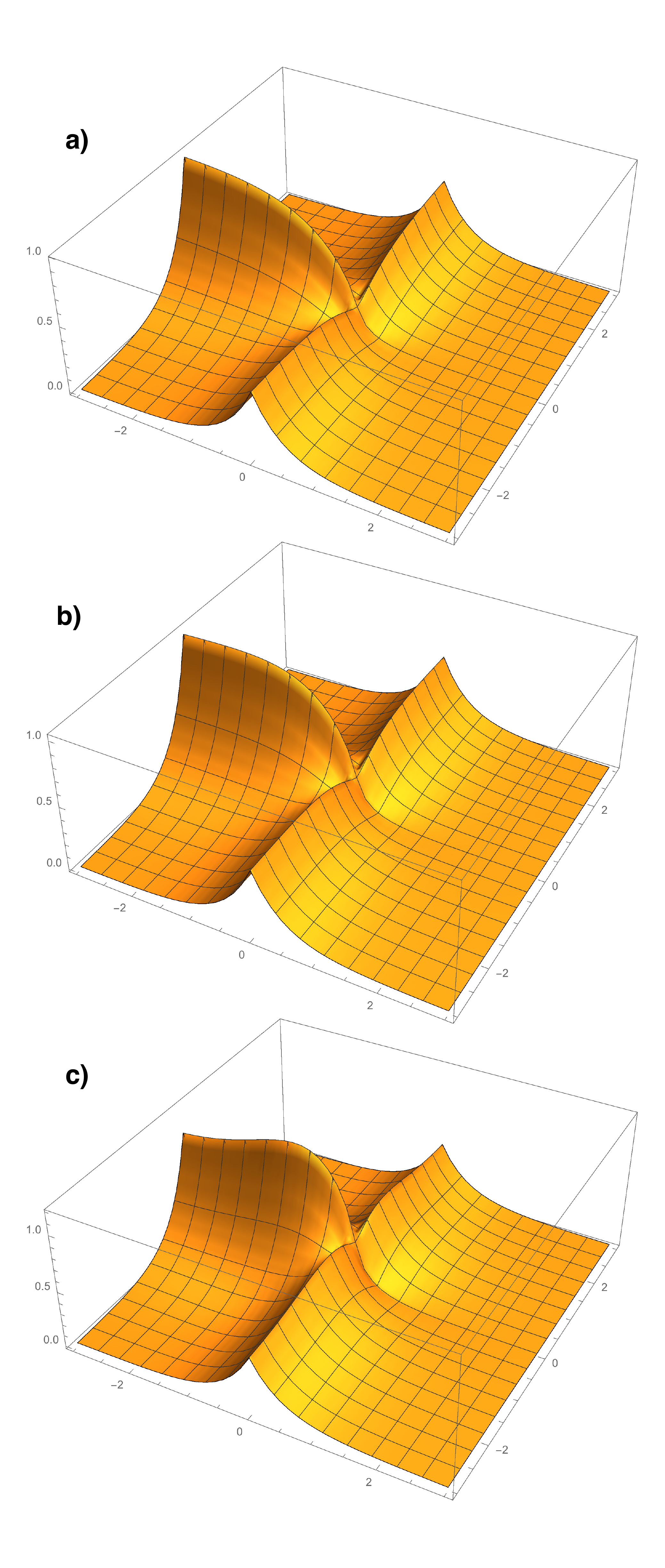}
\caption{Total density $|\Psi_{E}|^{2}$ as a function of $x$ and $y$ in units of $1/U$, for $E/U=0.1,0.5,0.95$ successively from top to bottom. These plots are first-order interpolations of 41-by-41 grids of $x,y$ values obtained by numerical integration of the integral representation. Plots which focus on the intersection at higher resolution confirm that the solutions are smoothly continuous. The sharp ridges of the maxima are not numerical artifacts; the ZLM solutions really are $\sim e^{-|U||x|}$ or $\sim e^{-|U||y|}$ .}
\end{figure}

The total density $|\Psi_{E}|^{2}$ is also shown in Fig.~5 as an intensity plot, with arrows showing the expectation values of the $x$ and $y$ components of the local spin. 
\begin{figure}[H]\label{greenfig}
\center\includegraphics[width=0.8\columnwidth]{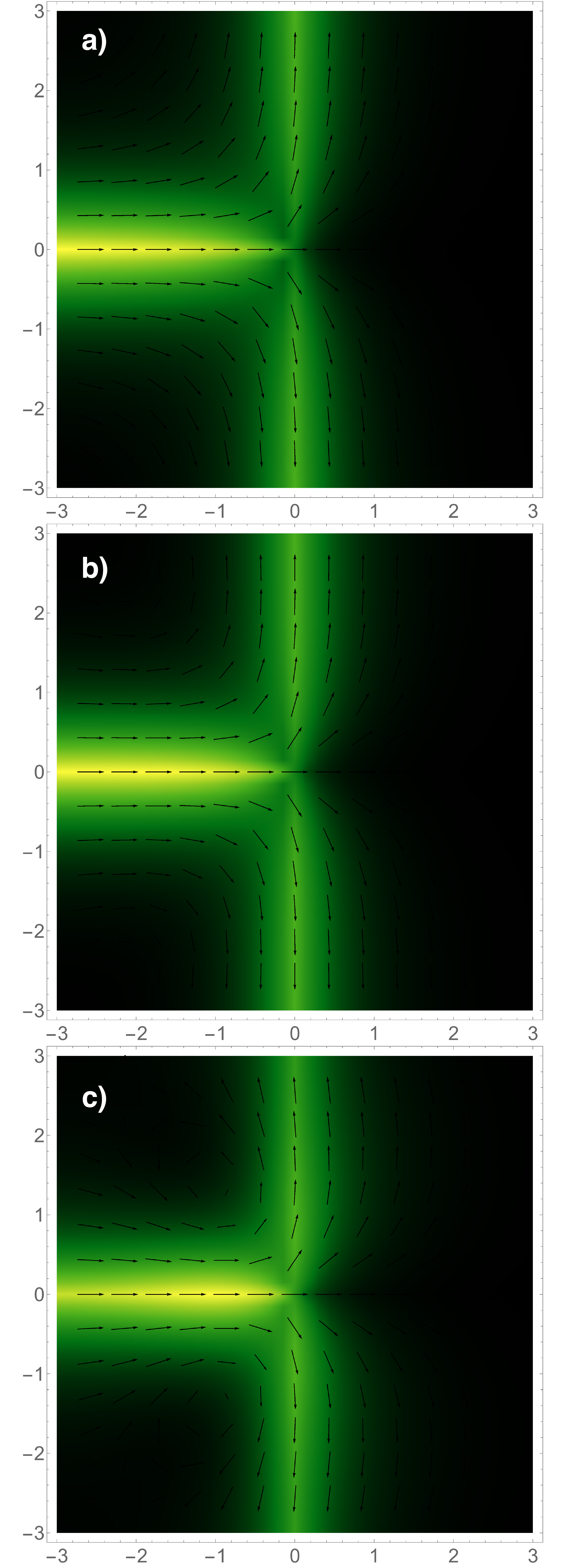}
\caption{Total density $|\Psi_{E}|^{2}$ (background brightness) and spin projection $(\langle\sigma_{x}\rangle,\langle\sigma_{y}\rangle)$ (arrow directions), as functions of $x$ and $y$ in units of $1/U$, for $E/U=0.1,0.5,0.95$ successively from top to bottom.}
\end{figure}

Another notable feature revealed by the plots is that the exact continuum eigenspinors make quite sharp `T's, even for $E$ within a few percent of $U$. This may be somewhat surprising; one could imagine instead that particles turning sharp corners might tend to veer further away from the zero-line axes, or that a stream of fluid encountering a junction might overflow its channel. Figures~4 and 5 show neither of those effects. Instead the quasi-one-dimensional confinement of the ZLMs persists robustly through the intersection, even when the binding energy $U-E$ is much smaller than $U$.

In final summary: checkerboard potentials act exactly as coherent beamsplitters for chiral two-dimensional Dirac zero-line modes. Even quite near the bulk energy gap edges, the charge profile of the in-gap eigenspinors is a remarkably clean `T'. The effective T-junction brings a phase shift whose energy dependence is nearly linear for $E<U\leq 1/2$.

\section{Discussion and Outlook}
\subsection{Asymmetrical intersections and corners}
In future work it may prove possible to extend the solution technique that we have presented to problems which lack our square symmetry. This would allow analytical description of general circuits in which contours of zero potential serve as `wires made of nothing', and the currents traveling through them occupy topologically protected edge states. The numerical lattice results of Ref.~\cite{Qiao2011,Qiao2014} have already indicated that zero-line modes (ZLMs) can turn corners, and split unequally when intersecting at angles other than ninety degrees. Analytical approaches to the Dirac equation have previously shown how ZLMs can follow curves, and how two nearby parallel ZLMs can influence each other by tunneling \cite{Tudorovskiy2012}.

\subsection{Interferometry}
ZLM circuitry with corners and intersections may even include SQUID-like interferometric devices made entirely of carbon \cite{Ji2003}. We have confirmed that a single ZLM wave packet made up of $\Psi_{E}$ will enter a topological intersection along the negative $x$-axis, and split coherently into two outgoing wave packets traveling in opposite directions along the $y$-axis. Superpositions of $\Psi_{E}$ and $\Psi'_{E}$ can contain waves incoming from both positive and negative ends of the $x$-axis, with a definite relative phase $\zeta$; the amplitudes of waves exiting along the positive and negative $y$-axes then depend on this relative phase. The simplest example is with equal-modulus superpositions of the form
\begin{widetext}\begin{eqnarray}\label{superposition}
\lefteqn{
\frac{1}{\sqrt{2}}\left[e^{i\zeta}\Psi_{E}(x,y) + e^{-i\zeta}\sigma_{y}\Psi_{E}(-x,y) \right]=}\nonumber\\
&&+ \theta(-x) \frac{e^{i\zeta}}{\sqrt{2}}e^{iEx}e^{-U|y|}\left(\begin{matrix}1\\ 1\end{matrix}\right) - i \theta(x) \frac{e^{-i\zeta}}{\sqrt{2}}e^{-iEx}e^{-U|y|}\left(\begin{matrix}1\\ -1\end{matrix}\right)\nonumber\\
&&+e^{-2i\alpha(E/U)}\left[\cos(\zeta)\theta(y)e^{-U|x|}e^{iEy}\left(\begin{matrix}\frac{1-i}{\sqrt{2}}\\ \frac{1+i}{\sqrt{2}}\end{matrix}\right)+i\sin(\zeta)\theta(-y)e^{-U|x|}e^{-iEy}\left(\begin{matrix}\frac{1+i}{\sqrt{2}}\\ \frac{1-i}{\sqrt{2}}\end{matrix}\right)\right]\nonumber\\
&&+\frac{e^{i\zeta}}{\sqrt{2}}\Delta\Psi_{E}(x,y)+\frac{e^{-i\zeta}}{\sqrt{2}}\sigma_{y}\Delta\Psi_{E}(-x,y)\;.
\end{eqnarray}\end{widetext}
The $\cos(\zeta)$ and $\sin(\zeta)$ terms in the third line here show that the relative phase $\zeta$ between the incident waves along the $x$-axis (terms in the second line) controls how much of the output goes in the positive or negative $y$-directions (third line terms). One can therefore imagine Mach-Zehnder interferometers based on topological intersections which coherently split and recombine ZLM wave packets to produce phase-dependent output currents in different output ports. A crucial question for such developments, however, is the effect of screened Coulomb interactions among real electrons.
 
\subsection{Mapping intersecting edge modes onto a one-dimensional model}
Another extension of this work would therefore consider intersecting ZLMs that are populated by fermions having short-ranged interactions. The ZLMs are quasi-one-dimensional, inasmuch as our solutions decay exponentially away from the $x$ and $y$ axes, and we have the two one-parameter families of orthogonal states which are expected for two one-dimensional lines, namely $\Psi_{E}$ and $\tilde{\Psi}_{E}$. We could therefore anticipate describing the interacting fermions in these ZLMs using Luttinger liquid theory. To do this will require an explicit mapping of the two-dimensional $\Psi_{E}(x,y)$ and $\tilde{\Psi}_{E}(x,y)$ onto a one-dimensional model; because of the splitting behavior at the intersection, however, this mapping is not quite trivial. Without actually analyzing any effects of interactions, we can set up this mapping in a way that will then be suitable for future studies of the interacting problem. It will turn out to present the topological intersection at $(x,y)=(0,0)$ as a sort of effective impurity which links two one-dimensional `wires' in a non-trivial way.

The interacting many-body problem is most easily treated in second quantization, where the Dirac spinor becomes a quantum field composed of eigenmodes with operator-valued coefficients that destroy fermions (or create holes). If we assume that temperatures and all other excitation energy scales are small compared to the bulk gap $U$, then only the $\Psi_{E}$ and $\tilde{\Psi}_{E}$ modes will be excited, and the quantum field is effectively
\begin{eqnarray}\label{fieldop}
\hat{\Psi}(x,y)=\int\!dE\,\left(\Psi_{E}(x,y)\hat{a}_{E\dashv}+\tilde{\Psi}_{E}(x,y)\hat{a}_{E\vdash}\right)\;.
\end{eqnarray}
Since $\tilde{\Psi}_{E}(x,y)=i\sigma_{y}\Psi_{E}(-x,y)$, we can consider that $\hat{a}^{\dagger}_{E\dashv}$ creates a fermion which is incident from the left (and then splits coherently into a superposition of being transmitted up and down) while $\hat{a}^{\dagger}_{E\vdash}$ creates one incident from the right (which then splits coherently into a similar superposition but with opposite relative phase between the two transmitted parts).

The fact that our $\Psi_{E}$ and $\tilde{\Psi}_{E}$ are solutions of the time-independent Dirac equation means that the single-particle energy is diagonal in these modes:
\begin{eqnarray}
	\hat{H}_{0}&=&\int\!dx\,dy\,\hat{\Psi}^{\dagger}\hat{\mathcal{H}}\hat{\Psi}\nonumber\\
	&=&\sum_{j=\dashv,\vdash}\int\!dE\,E\,\hat{a}^{\dagger}_{Ej}\hat{a}_{Ej}\;.
\end{eqnarray}
If there are no interactions among the particles, the many-body problem is thus fully solved already; interactions, however, bring the usual problem that the interaction Hamiltonian and single-particle Hamiltonian cannot be diagonalized simultaneously. Short-ranged screened Coulomb interactions among dressed electrons can typically be described with an interaction Hamiltonian density proportional to the square of the local charge density,
\begin{eqnarray}\label{Hint}
	\hat{H}_{int} = \int\!d^{2}r\,d^{2}r'\,\hat{\Psi}^{\dagger}(\mathbf{r})\hat{\Psi}(\mathbf{r})V(\mathbf{r}-\mathbf{r}')\hat{\Psi}^{\dagger}(\mathbf{r}')\hat{\Psi}(\mathbf{r}')
\end{eqnarray}
(up to appropriate normal-ordering), for a screened Coulomb potential $V(\mathbf{r})$. Inserting (\ref{fieldop}) and integrating reveals $\hat{H}_{int}$ to be a non-diagonal term involving products of four creation or destruction operators.

Since we are neglecting excitations with energies outside the bulk gap $|E|<U$, however, our field operator $\hat{\Psi}$ according to (\ref{fieldop}) is actually restricted quite closely to the $x-$ and $y-$axes. To be precise, if we define the one-dimensional quantum fields
\begin{eqnarray}\label{1dDef}
	\hat{\Phi}_{j=\dashv,\vdash}(z) = \frac{1}{\sqrt{2\pi}}\int\!dE\,e^{iEz}\hat{a}_{Ej}\;,
\end{eqnarray}
then simply inserting our solution (\ref{psiE2}) for $\Psi_{E}(x,y)$ and $\tilde{\Psi}_{E}(x,y)=i\sigma_{y}(-x,y)$ into the definition of the field operator (\ref{fieldop}) reveals that
\begin{eqnarray}\label{TheReal55}
	\hat{\Psi}(x,y)&\doteq& \theta(-x) \left(\begin{matrix}1\\ 1\end{matrix}\right)e^{-U|y|}\hat{\Phi}_{\dashv}(x)\\
&&+\theta(x) \left(\begin{matrix}1\\ -1\end{matrix}\right)e^{-U|y|}\hat{\Phi}_{\vdash}(-x)\nonumber\\
&&+\theta(y)\left(\begin{matrix}\frac{1-i}{\sqrt{2}}\\ \frac{1+i}{\sqrt{2}}\end{matrix}\right)e^{-U|x|}\frac{\hat{\Phi}_{\dashv}(y)+\hat{\Phi}_{\vdash}(y)}{\sqrt{2}}\nonumber\\
&&+\theta(-y)\left(\begin{matrix}\frac{1+i}{\sqrt{2}}\\ \frac{1-i}{\sqrt{2}}\end{matrix}\right)e^{-U|x|}\frac{\hat{\Phi}_{\dashv}(-y)-\hat{\Phi}_{\vdash}(-y)}{\sqrt{2}}\nonumber\\
&&+\Delta\Psi_{0}(x,y)\hat{\Phi}_{\dashv}(0)+i\sigma_{y}\Delta\Psi_{0}(-x,y)\hat{\Phi}_{\vdash}(0)\;.\nonumber
\end{eqnarray}
where the approximation indicated by the $\doteq$ is only that of replacing $\Delta\Psi_{E}\doteq\Delta\Psi_{0}$. We have seen that this approximation is excellent except for $E$ very close to the top or bottom of the gap. 

The fact that we can thus express all the in-gap modes of the two-dimensional $\hat{\Psi}(x,y)$ in terms of the two one-dimensional fields $\hat{\Phi}_{\dashv,\vdash}(\pm x)$ and $\hat{\Phi}_{\dashv,\vdash}(\pm y)$ means that the two-dimensional integral in $\hat{H}_{int}$ can in fact be reduced to sums of one-dimensional integrals involving the one-dimensional fields. This is a valuable insight, because even though $\hat{H}_{int}$ and $\hat{H}_{0}$ still do not commute, interacting fermions in one dimension are amenable to powerful analytical techniques based on so-called \textit{bosonization}. Moreover, the single-particle Hamiltonian can also be expressed very simply in terms of $\hat{\Phi}_{\dashv,\vdash}(z)$, because the orthogonality of plane waves implies that
\begin{eqnarray}
	\sum_{j=\dashv,\vdash}\int\!dz\,\hat{\Phi}_{j}^{\dagger}(z)(-i\partial_{z})\hat{\Phi}_{j}(z)&\equiv&\sum_{j=\dashv,\vdash}\int\!dE\,E\,\hat{a}^{\dagger}_{Ej}\hat{a}_{Ej}\nonumber\\
	&=&\hat{H}_{0}\;.
\end{eqnarray}
The prospects therefore seem good for mapping the full Hamiltonian $\hat{H}=\hat{H}_{0}+\hat{H}_{int}$ onto that of a fermionic field theory in one dimension.

A difficulty presents itself, however: when we perform the $x$ and $y$ integrals in (\ref{Hint}) using the expression (\ref{TheReal55}) for $\hat{\Psi}$ in terms of $\hat{\Phi}_{\dashv,\vdash}$, then even for the idealized limit of the short-ranged screened $V(\mathbf{r})$ as a contact interaction, we obtain
\begin{eqnarray}\label{Hint2}
	\hat{H}_{int} &=& G\int_{-\infty}^{0}\!dz\,\left[(\hat{\Phi}_{\dashv}^{\dagger}\hat{\Phi}_{\dashv})^{2}+(\hat{\Phi}_{\vdash}^{\dagger}\hat{\Phi}_{\vdash})^{2}\right]\nonumber\\
&&+ G\int_{0}^{\infty}\!dz\,\left[\left(\frac{(\hat{\Phi}_{\dashv}^{\dagger}+\hat{\Phi}^{\dagger}_{\vdash})(\hat{\Phi}_{\dashv}+\hat{\Phi}_{\vdash})}{2}\right)^{2}\right.\nonumber\\
&&\qquad +\left. \left(\frac{(\hat{\Phi}_{\dashv}^{\dagger}-\hat{\Phi}^{\dagger}_{\vdash})(\hat{\Phi}_{\dashv}-\hat{\Phi}_{\vdash})}{2}\right)^{2}\right]\nonumber\\
&&+\sum_{jklm}G_{jklm}\hat{\Phi}^{\dagger}_{j}(0)\hat{\Phi}_{k}(0)\hat{\Phi}^{\dagger}_{l}(0)\hat{\Phi}_{m}(0)\nonumber\\
&=& G\int_{-\infty}^{0}\!dz\,\left[(\hat{\Phi}_{\dashv}^{\dagger}\hat{\Phi}_{\dashv})^{2}+(\hat{\Phi}_{\vdash}^{\dagger}\hat{\Phi}_{\vdash})^{2}\right]\nonumber\\
&&+ \frac{G}{2}\int_{0}^{\infty}\!dz\,\left[(\hat{\Phi}_{\dashv}^{\dagger}\hat{\Phi}_{\dashv}+\hat{\Phi}_{\vdash}^{\dagger}\hat{\Phi}_{\vdash})^{2}\right.\nonumber\\
&&\qquad +\left. 2 \hat{\Phi}_{\dashv}^{\dagger}\hat{\Phi}_{\dashv}\hat{\Phi}_{\vdash}^{\dagger}\hat{\Phi}_{\vdash}+\hat{\Phi}_{\dashv}^{\dagger2}\hat{\Phi}_{\vdash}^{2}+\hat{\Phi}_{\vdash}^{\dagger2}\hat{\Phi}_{\dashv}^{2}\right]\nonumber\\
&&+\sum_{jklm}G_{jklm}\hat{\Phi}^{\dagger}_{j}(0)\hat{\Phi}_{k}(0)\hat{\Phi}^{\dagger}_{l}(0)\hat{\Phi}_{m}(0)
\end{eqnarray}
where $G$ and $G_{jklm}$ are effective constants defined from the microscopic interaction $V(\mathbf{r})$ by the two-dimensional integration. The $G_{jklm}$ term is an effective two-body impurity term at $z=0$; it is present because the intersecting eigenmodes have the localized $\Delta\Psi_{0}$ form near the intersection which is different from their asymptotic forms along the $x$ and $y$ axes, and so the interaction energy is somewhat different in this region.

The difficulty which (\ref{Hint2}) presents is that the effective interaction Hamiltonian density mixes the fields $\hat{\Phi}_{\dashv}$ and $\hat{\Phi}_{\vdash}$ for $z>0$. This curious feature in the one-dimensional effective model for the intersecting ZLMs is an essential consequence of the coherent splitting at the topological intersection: since $\hat{\Phi}_{\dashv}$ is composed of second-quantized operators $\hat{a}_{E\dashv}$ which destroy particles in $\Psi_{E}$ states, and since $\Psi_{E}(x,y)$ consists of an incident wave on the negative $x$-axis plus an equal superposition of outgoing waves in both directions on the $y$-axis, $\hat{\Phi}_{\dashv}$ can effectively destroy particles locally along the negative $x$-axis, but it does nothing to any particles on the positive $x$-axis, and what it does to particles on the $y$-axis is a coherent superposition of destroying a particle at $y$ and $-y$. Conversely, $\hat{\Phi}_{\vdash}$ can destroy particles on the positive $x$-axis, does not affect particles on the negative $x$-axis, and destroys particles on the $y$-axis at a superposition of $y$ and $-y$, opposite in phase to the action of $\hat{\Phi}_{\dashv}$.

The $x$-axis is not really any more local than the $y$-axis, however. We could simply perform a basis rotation in the $j=1,2$ indices, to construct $\hat{\Phi}_{\pm} = (\hat{\Phi}_{\dashv}\pm \hat{\Phi}_{\vdash})/\sqrt{2}$. This orthogonal transformation would leave $\hat{H}_{0}$ unchanged in form, and while it would diagonalize the $z>0$ part of $\hat{H}_{int}$ in field index-space, the $z<0$ part of $\hat{H}_{int}$ would become non-diagonal in the $\pm$ indices in precisely the same way that the $z>0$ part is non-diagonal in the $j=\dashv,\vdash$ indices. There is no global basis rotation which can diagonalize $\hat{H}_{int}$ everywhere at once. 

This is unfortunate, because non-diagonal interaction terms like $\hat{\Phi}_{\dashv}^{\dagger 2}\hat{\Phi}_{\vdash}^{2}$ are a significant inconvenience in bosonization techniques for dealing with interacting fermions in one dimension. As well as the mathematical complications, there are also challenges to physical intuition. One cannot quite think of the $\dashv$ and $\vdash$ fields as existing on two distinct wires, because for $z>0$ the wires are rotated into coherent superpositions of each other. What we can therefore do is to introduce a $z$-\textit{dependent} basis rotation:
\begin{eqnarray}\label{Zdef}
	\left(\begin{matrix}\hat{Z}_{1}(z)\\ \hat{Z}_{2}(z) \end{matrix}\right) = e^{i\frac{\pi}{4}\sigma_{y}\theta(z)}\left(\begin{matrix}\hat{\Phi}_{\dashv}(z)\\ \hat{\Phi}_{\vdash}(z) \end{matrix}\right)\;.
\end{eqnarray}
This transformation provides $\hat{Z}_{1}=\hat{\Phi}_{\dashv}$ for $z<0$ but $\hat{Z}_{1}=\hat{\Phi}_{+}$ for $z>0$ --- and correspondingly for $\hat{Z}_{2}$. In other words, the ZLM along the negative $x$-axis is now represented by $\hat{Z}_{1}(z)$ for $z<0$, just as it was by $\hat{\Phi}_{\dashv}$, but where the ZLM along the positive $y$-axis corresponded to the superposition $(\hat{\Phi}_{\dashv}+\hat{\Phi}_{\vdash})/\sqrt{2}$, it now corresponds simply to the same single field $\hat{Z}_{1}(z)$, for $z>0$. The ZLMs along the positive $x$-axis and the negative $y$-axis are similarly represented by $\hat{Z}_{2}(z)$, for negative and positive $z$, respectively. Each half-axis ZLM is represented locally by a single one-dimensional quantum field.

We thereby diagonalize $\hat{H}_{int}$ (except perhaps for the $G_{jklm}$ impurity term at $z=0$), obtaining
\begin{eqnarray}
	\hat{H}_{int} &=& \sum_{j=1,2}\int\! dz\,(\hat{Z}_{j}^{\dagger}\hat{Z}_{j})^{2} \nonumber\\
&&+ \sum_{jklm}\Gamma_{jklm}\hat{Z}^{\dagger}_{j}(0)\hat{Z}^{\dagger}_{k}(0)\hat{Z}_{l}(0)\hat{Z}_{m}(0)\;,
\end{eqnarray}
where $\Gamma_{jklm}$ is $G_{jklm}$ re-expressed in the new $j$ basis. One can consider the interactions in this system as being like those among particles in two identical but separate one-dimensional wires, which cross at a common point whose position is labeled as $z=0$ in both wires. 

The fact that effective wire 1 extends along the physical negative $x-$ and positive $y-$axes, while wire 2 spans the other two half-axes, does not really represent any preference for left-handed angles in the checkerboard problem. A similar transformation to (\ref{Zdef}) but with opposite signs could have represented the negative $x-$ and $y-$axes as wire 1 and the two positive axes as wire 2. The two alternative mappings of ZLMs onto effective one-dimensional fields are entirely equivalent---if we remember to include the following point.

In addition to conveniently making $\hat{H}_{int}$ local in $z$, the $z$-dependent transformation from $\hat{\Phi}_{\dashv,\vdash}$ to $\hat{Z}_{1,2}$ also does something else. It adds an impurity term to the single-particle energy:
\begin{eqnarray}
	\hat{H}_{0}&=& \sum_{j=\dashv,\vdash}\int\!dz\,\hat{\Phi}_{j}^{\dagger}(z)(-i\partial_{z})\hat{\Phi}_{j}(z)\nonumber\\
	&\equiv&\sum_{j=1,2}\int\!dz\,\hat{Z}_{j}^{\dagger}(z)(-i\partial_{z})\hat{Z}_{j}(z)\nonumber\\
&&+iJ[\hat{Z}^{\dagger}_{1}(0)\hat{Z}_{2}(0)-\hat{Z}^{\dagger}_{2}(0)\hat{Z}_{1}(0)]
\end{eqnarray}
for $J=\pi/4$. 
In terms of the effective one-dimensional model, therefore, the topological intersection behaves essentially as a localized impurity transferring particles between the two one-dimensional channels at their intersection point $z=0$. In this respect it is not qualitatively different from similar contact points between other one-dimensional conduits, such as crossed nanotubes. And although in this paper we have only treated the perfect square checkerboard potential which leads to $J=\pi/4$, we can anticipate that intersections at other angles, or imperfect intersections in which two zero-potential contours approach each other but fail to cross, may be represented with a similar impurity term to the above, with some other values of $J$. We thereby provide support for the prospects suggested in \cite{Qiao2011,Qiao2014} of realizing intersections between real one-dimensional systems as intersections of topologically protected zero-line modes in designer potentials.

\bibliography{references}

\end{document}